\documentclass[twocolumn]{aastex631}

\shorttitle{ALMA detection of [OIII] 88\,$\rm \mu m$ at $z=12$}
\shortauthors{Zavala, Bakx, Mitsuhashi et al.}
\begin{document}

\title{ALMA detection of [OIII] 88\,\boldmath$\mu\rm m$ at \boldmath$z=12.33$: \\
Exploring the Nature and Evolution of GHZ2 as a Massive Compact Stellar System}

\author[0000-0002-7051-1100]{Jorge A. Zavala}
\affiliation{National Astronomical Observatory of Japan, 2-21-1 Osawa, Mitaka, Tokyo 181-8588, Japan}

\author{Tom Bakx}
\affiliation{Department of Space, Earth, \& Environment, Chalmers University of Technology, Chalmersplatsen 4 412 96 Gothenburg, Sweden }

\author{Ikki Mitsuhashi}
\affiliation{Waseda Research Institute for Science and Engineering, Faculty of Science and Engineering, Waseda University, 3-4-1, Okubo, Shinjuku, Tokyo 169-8555, Japan }

\author{Marco Castellano}
\affiliation{INAF - Osservatorio Astronomico di Roma, via di Frascati 33, 00078 Monte Porzio Catone, Italy }

\author{Antonello Calabro}
\affiliation{INAF - Osservatorio Astronomico di Roma, via di Frascati 33, 00078 Monte Porzio Catone, Italy }

\author{Hollis Akins}
\affiliation{ Department of Astronomy, The University of Texas at Austin, 2515 Speedway Boulevard Stop C1400, Austin, TX 78712, USA}

\author{Veronique Buat}
\affiliation{ Aix Marseille Univ, CNRS, CNES, LAM, Marseille, France}

\author{Caitlin M. Casey}
\affiliation{ Department of Astronomy, The University of Texas at Austin, 2515 Speedway Boulevard Stop C1400, Austin, TX 78712, USA}

\author{David Fernandez-Arenas}
\affiliation{Canada-France-Hawaii Telescope, Kamuela, HI 96743, USA}
\affiliation{Instituto de Radioastronom{\'i}a y Astrof{\'i}sica, UNAM Campus Morelia, Apartado postal 3-72, 58090 Morelia, Michoac{\'a}n, Mexico}

\author{Maximilien Franco}
\affiliation{ Department of Astronomy, The University of Texas at Austin, 2515 Speedway Boulevard Stop C1400, Austin, TX 78712, USA}

\author[0000-0003-3820-2823]{Adriano Fontana}
\affiliation{INAF - Osservatorio Astronomico di Roma, Via Frascati 33, 00078, Monte Porzio Catone, Italy }

\author{Bunyo Hatsukade}
\affiliation{National Astronomical Observatory of Japan, 2-21-1 Osawa, Mitaka, Tokyo 181-8588, Japan}
\affiliation{Graduate Institute for Advanced Studies, SOKENDAI, Osawa, Mitaka, Tokyo 181-8588, Japan}
\affiliation{Institute of Astronomy, Graduate School of Science, The University of Tokyo, 2-21-1 Osawa, Mitaka, Tokyo 181-0015, Japan}

\author[0000-0001-6947-5846]{Luis C. Ho}
\affil{Kavli Institute for Astronomy and Astrophysics, Peking University, Beijing 100871, China}
\affil{Department of Astronomy, School of Physics, Peking University, Beijing 100871, China}

\author{Ryota Ikeda}
\affiliation{National Astronomical Observatory of Japan, 2-21-1 Osawa, Mitaka, Tokyo 181-8588, Japan}
\affiliation{Department of Astronomy, School of Science, SOKENDAI (The Graduate University for Advanced Studies), 2-21-1 Osawa, Mitaka, Tokyo 181-8588, Japan}

\author{Jeyhan Kartaltepe}
\affiliation{Laboratory for Multiwavelength Astrophysics, School of Physics and Astronomy, Rochester Institute of Technology, 84 Lomb Memorial Drive, Rochester, NY 14623, USA}

\author[0000-0002-6610-2048]{Anton M. Koekemoer}
\affiliation{Space Telescope Science Institute, 3700 San Martin Drive, Baltimore, MD 21218, USA}

\author{Jed McKinney}
\affiliation{Department of Astronomy, The University of Texas at Austin, 2515 Speedway Boulevard Stop C1400, Austin, TX 78712, USA}

\author[0000-0002-8951-4408]{Lorenzo Napolitano}
\affiliation{INAF – Osservatorio Astronomico di Roma, via Frascati 33, 00078, Monteporzio Catone, Italy}
\affiliation{Dipartimento di Fisica, Università di Roma Sapienza, Città Universitaria di Roma - Sapienza, Piazzale Aldo Moro, 2, 00185, Roma, Italy}

\author{Pablo G. P{\'e}rez-Gonz{\'a}lez}
\affiliation{Centro de Astrobiología (CAB), CSIC-INTA, Ctra. de Ajalvir km 4, Torrejón de Ardoz, E-28850, Madrid, Spain}

\author{Paola Santini}
\affiliation{INAF - Osservatorio Astronomico di Roma, via di Frascati 33, 00078 Monte Porzio Catone, Italy}

\author{Stephen Serjeant}
\affiliation{School of Physical Sciences, The Open University, Walton Hall, Milton Keynes, MK7 6AA }

\author{Elena Terlevich}
\affiliation{Instituto Nacional de Astrof\'\i sica, \'Optica y Electr\'onica,Tonantzintla, Puebla, M\'exico}
\affiliation{Institute of Astronomy, University of Cambridge, Cambridge, CB3 0HA, UK}
\affiliation{Facultad de Astronomía y Geofísica, Universidad de La Plata, La Plata, Argentina}

\author{Roberto Terlevich}
\affiliation{Instituto Nacional de Astrof\'\i sica, \'Optica y Electr\'onica,Tonantzintla, Puebla, M\'exico}
\affiliation{Institute of Astronomy, University of Cambridge, Cambridge, CB3 0HA, UK}
\affiliation{Facultad de Astronomía y Geofísica, Universidad de La Plata, La Plata, Argentina}

\author[0000-0003-3466-035X]{{L. Y. Aaron} {Yung}}
\affiliation{Space Telescope Science Institute, 3700 San Martin Drive, Baltimore, MD 21218, USA}

% \collaboration{20}{(AAS Journals Data Editors)}

% \author{Amy Hendrickson}
% \altaffiliation{AASTeX v6+ programmer}
% \affiliation{TeXnology Inc.}

% \author{Julie Steffen}
% \affiliation{AAS Director of Publishing}
% \affiliation{American Astronomical Society \\
% 1667 K Street NW, Suite 800 \\
% Washington, DC 20006, USA}



\begin{abstract}
We present ALMA observations on the high-redshift galaxy GHZ2 and report a successful detection of the rest-frame 88\,$\mu\rm m$ atomic transition from doubly-ionized Oxygen at $z=12.3327\pm0.0005$. Based on these observations, combined with additional constraints on the [OIII] 52\,$\mu\rm m$ line luminosity and previous JWST data, we argue that GHZ2 is likely powered by compact and young star formation, and show that it follows well-established relationships found for giant HII regions and metal-poor star-forming dwarf galaxies that are known to host bright super star clusters. Additionally, these observations provide new constraints on the Oxygen electron density ($100\lesssim n_e\,{\rm[cm^{-3}]}\lesssim4,000$) and dynamical mass ($M_{\rm dyn}\approx3-8\times10^8\rm\,M_\odot$).
The existence of these massive starburst systems 13.3\,Gyr ago might explain the origin of today's globular clusters, a long-standing question in astronomy. To test this, we present observational probes to investigate whether sources like GHZ2 are linked to the formation of today's globular clusters or other more massive compact stellar systems. 

\end{abstract}



\section{Introduction} \label{sec:intro}
The discovery of  bright galaxies at very high redshifts ($z\gtrsim10$) by the James Webb Space Telescope (JWST) is challenging  our  understanding of galaxy formation and evolution. Their high luminosities ($M_{UV}<-20$) and inferred volume density are in tension with the predictions from widely-adopted (pre-JWST) theoretical studies (e.g. \citealt{Adams2024,Castellano2023,Finkelstein2023,Harikane2024b}). This could be the result of very young stellar populations and/or  bursty star formation, enhanced star formation efficiencies in the early Universe, contribution from active black holes, or strong evolution in the dust properties, among others, including a non-universal initial mass function or modifications to the standard cosmology itself (e.g. \citealt{Boylan-Kolchin2023,Feldmann2024,Ferrara2023,Liu2022,Sun2023,Yung2024}). 

These galaxies provide thus crucial 
insights into the earliest phases of galaxy formation within the first few hundred million years after the Big Bang, as well as into the galaxy-dark matter connection and black hole-galaxy co-evolution. To date, more than ten galaxies have been spectroscopically confirmed at $z>10$ (see recent compilation by \citealt{Roberts-Borsani2024}), reaching up to $z\sim14$ \citep{Carniani2024}, but their nature and primary source of excitation remain unclear, especially for the brightest and most compact systems. {\it Are they primarily powered by intense star formation or do they host active galactic nuclei (AGN) that contribute significantly to their luminosity?}

To address this question, spectroscopic observations to probe the physics of their multiphase interstellar media are essential. At high redshifts, the JWST NIRSpec instrument has became one of the workhorses for characterizing these systems thanks to its high sensitivity and wavelength coverage (e.g. \citealt{ArrabalHaro2023,Bunker2023,Curtis-Lake2023,Fujimoto2023,Sanders2023,Wang2023}), but at $z\gtrsim9-10$, it provides access mainly to rest-frame UV lines. Recently, the Mid-Infrared Instrument (MIRI) has also been used to study some of the highest redshift galaxies since it probes the well studied (and typically bright) rest-frame optical emission lines up to $z\gtrsim 14-15$, offering valuable complementary data to the NIRSpec observations (e.g. \citealt{Alvarez-Marquez2024,Hsiao2024,Zavala2024}). 
Furthermore, the Atacama Large Millimeter/submillimeter Array (ALMA)  is uniquely equipped to characterize high-redshift galaxies through the detection of far-infrared emission lines, which can also serve as powerful diagnostics for the physical conditions of distant galaxies (e.g. \citealt{Inoue2014,Inoue2016,Bakx2020,Carniani2017,Hashimoto2019}), specifically  when using  multi-wavelength line diagnostics (e.g. \citealt{Katz2019,Nakazato2023}).

Nevertheless, some recent ALMA follow-ups on ultra-high-redshift galaxy candidates have resulted in non-detections (\citealt{Bakx2023,Kaasinen2023,Yoon2023}), preventing us from fully characterizing the nature of these systems and calling into question the ALMA's capabilities to study the $z\gtrsim10$ universe. Most of these observations, however, suffered from targets with highly uncertain redshifts (most of them limited to photometric redshifts), which require line-searches through multi-setup spectral scans, limiting the sensitivity of the observations and preventing the derivation of tight constraints on their physical properties.

In this study, we present deep ALMA observations targeting a spectroscopically confirmed galaxy within the Universe's first 400\,Myrs, GHZ2/GLASSz12  at $z=12.3$ (hereafter GHZ2; \citealt{Castellano2024,Zavala2024}; see also \citealt{Naidu2022,Castellano2022,Calabro2024}), and report a  successful detection of the [OIII] 88$\mu\rm m$  far-infrared emission line. This breakthrough allows us to place new constraints on the physical properties of this unique galaxy, which stands out as one of the brightest known objects at this epoch ($M_{UV}\approx-20.5$). Moreover, this study  provides critical insights into the nature and evolution of the luminous high-redshift galaxies recently discovered by JWST and the mechanisms powering their extraordinary brightness, while demonstrating  the feasibility of studying these early systems with ALMA. 

The enormous potential of the ALMA-JWST synergy is further demonstrated by \citet{Carniani2024b} and \citet{Schouws2024}, who reported subsequent ALMA observations on a higher redshift galaxy, JADES-z14-0, with a successful detection of the [OIII] 88$\mu\rm m$ transition at $z=14.18$.

This paper is structured as follows. Section \ref{secc:observations} presents the details of the ALMA observations and data reduction, as well as the identification of the [OIII] 88$\micron$ line. The main results of the paper and the comparison of the properties of GHZ2 and those from local and low-redshift star-forming systems are presented in Section \ref{secc:results}. In Section \ref{secc:discussion} we discuss circumstantial evidence relating GHZ2, and other bright high-redshift galaxies, to globular clusters and other compact stellar systems. Finally, a summary of this work is presented in Section \ref{secc:summary}.
A companion paper (Mitsuhashi et al. in prep.) will present the dust constraints inferred from the ALMA continuum observations (which remains undetected at $>3\sigma$). \\

Throughout this paper, we assume a flat $\Lambda$CDM cosmology with $\Omega_\mathrm{m} = 0.29$, $\Omega_\mathrm{\Lambda} = 0.71$ and $\rm H_0=69.6\,\rm km\,s^{-1}\,Mpc^{-1}$.

\section{ALMA observations, data reduction, and data analysis}\label{secc:observations} 
\subsection{Observations and data reduction}

\begin{figure*}[t]
    \centering
    \includegraphics[width=\textwidth]{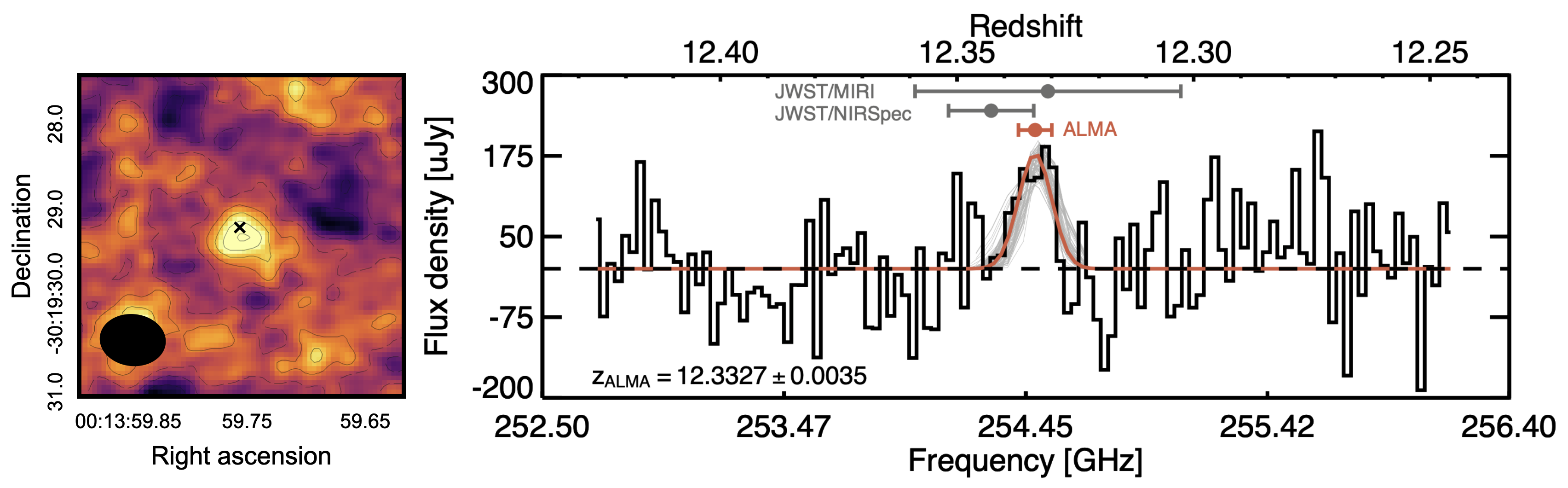}
    \caption{{\it Detection of the [OIII] 88\,$\mu\rm m$ transition at $z=12.3$}. The left panel shows the $3\farcs5\times3\farcs5$ moment-0 map of the line, along with the ALMA beam-size (black ellipse) and the JWST NIRCam position of GHZ2 (black cross). Contours show signal-to-noise ratio in steps of $\pm1\sigma$ (with dashed lines for negative values).  The $\sim5\sigma$ detection from ALMA is in very good agreement with the JWST position. The extracted spectrum from the peak pixel is shown on the right panel along with the best-fit Gaussian function (red solid line) and associated uncertainty (gray lines; see main text for details), implying a spectroscopic redshift of $z=12.3327\pm0.0035$. On the top, we show the associated redshifts inferred from the different instruments, including NIRSpec, MIRI, and ALMA. Thanks to the higher spectral resolution of the ALMA observations, the redshift precision is increased by $\sim5\times$. }
    \label{fig:ALMA_spectrum}
\end{figure*}

The source GHZ2 was previously targeted by ALMA using a Band 6 spectral scan strategy to search for the [OIII] 88$\,\mu\rm m$ line ($\nu_{\rm rest} = 3393.0062\,\rm GHz$) across the frequency range  234-263\,GHz, corresponding to $z\approx11.9-13.55$. Nevertheless, no robust detection was identified at the JWST position of the target (\citealt{Bakx2023}; see also \citealt{Popping2023}). After the spectroscopic redshift  confirmation by JWST/NIRSpec and JWST/MIRI \citep{Castellano2024,Zavala2024}, deeper ALMA observations were obtained in April 2024 around the refined expected frequency of the line at 254.45\,GHz as part of the DDT project 2023.A.00017.S (PI: J. Zavala). The new observations consist of $\sim4.9\,\rm h$ on-source ($\sim8.7\,\rm h$ including calibrations and overheads), which adds to $\sim7\,\rm h$ on-source in total when combining the cycle 9 and cycle 10 data-sets. Unfortunately, the new data do not cover the previous tentative line reported by \citet{Bakx2023} at $\nu=258.67\,\rm GHz$.

Additionally, Band 8 observations covering the redshifted [OIII] 52$\,\mu\rm m$ line ($\nu_{\rm rest} = 5785.8796\,\rm GHz$)  were also carried out as part of project 2023.A.00017.S. The spectral setup covers the frequency ranges $418.6-422.4\rm\, GHz$ and $430.8-434.5\rm\, GHz$, with an on-source time of $2.5\,\rm h$ (4.8\,\rm h including calibrations and overheads). These observations are by design shallower than the Band 6 data since were only aimed to provide a limit on the [OIII] 52$\,\mu\rm m$ in the case of an extreme electron density (see Section \ref{secc:electron_density}). Matching the Band 6 sensitivity would have required a much larger time investment that is beyond the scope of a DDT proposal.

Data reduction was performed as described in \cite{Bakx2023}, using CASA and following the standard ALMA pipeline workflow. Then, as part of our imaging process, we created data cubes from the $uv$-visibilties using a Natural weighting to maximize the signal-to-noise ratio (at the cost of angular resolution). For the Band 6 observations, during this process, we tested several choices for the pixel size (namely $0\farcs05$, $0\farcs075$, and $0\farcs10$) and channel width (35, 50, 75, and 100\,km\,s$^{-1}$), as well as the effect of slightly changing the phase center within a beam-size and slightly shifting the velocity of the first channel.  The typical sensitivity (r.m.s.) of these cubes is equivalent to $1\sigma=75\,\mu$Jy\,beam$^{-1}$ in 35\,km\,s$^{-1}$ channels, and have a typical beam size of $0\farcs36\times0\farcs28$. As described below, all of these datacubes are used to robustly estimate the uncertainties on the line parameters.

In the case of the Band 8 observations, the achieved $1\sigma$ r.m.s.  is around 500\,$\mu$Jy\,beam$^{-1}$ in 35\,km\,s$^{-1}$ channels, with a beam size of $0\farcs36\times0\farcs32$.

\subsection{The [OIII]88$\micron$ transition at $z = 12.3$: line properties}

We use the collection of datacubes described above to search for the  [OIII] 88$\micron$ emission line using an iterative process as follows. First, we extract the spectrum from each datacube exactly at the JWST position of GHZ2. Second, we identify the brightest channel within the expected frequency range of the redshifted [OIII]88$\micron$ line base on the JWST/MIRI and JWST/NIRSpec spectroscopic redshifts, and perform a Gaussian fit with $A,\sigma, \nu_{\rm cent}$ as free   parameters (with a flat zero continuum). Then, to improve the SNR of each extracted spectrum, we create a moment-0 map by collapsing the datacube within the full-width half-maximum (FWHM) of the best-fit Gaussian function and re-extract the spectrum but now at the position of the peak pixel in this moment-0 map. Finally, we perform a new Gaussian fit around the expected frequency and save the best-fit parameters. Reassuringly, all the extracted spectra (those extracted at the JWST position and at the peak ALMA pixel) identify the same feature as an emission line with a significance of $\approx4.5-5.2\sigma$, confirming the tentative detection reported in \citet{Zavala2024}. 

Figure \ref{fig:ALMA_spectrum} shows one of the extracted spectrum and associated moment-0 map (right and left panels, respectively) along with the adopted best-fit Gaussian function (red solid line). The best-fit Gaussian functions from other spectra are also shown to illustrate the typical variance in the data (gray lines). The fiducial
line properties and associated uncertainties are assumed to be the median values of all the different best-fit functions and their standard deviation. This approach allows us to properly take into account the impact of the imaging parameters in the shape of the line and to robustly estimate the uncertainties in the best-fit values. These best-fit properties are summarized in Table \ref{table:line_properties}.

\begin{deluxetable}{cc}[h]
\label{table:line_properties}
%% This is the title of the table.
\tablecaption{[OIII] 88$\,\mu\rm m$ line properties (uncorrected for gravitational amplification).}
\tablehead{\multicolumn{2}{c}{GHZ2}\\
\multicolumn{2}{c}{${\rm RA}=00$:13:59.76; $\rm Dec=-30$:19:29.16}} 
%% All data must appear between the \startdata and \enddata commands
\startdata
$z_{\rm [OIII]88\mu m} $  & $12.3327\pm 0.0035$ \\
$z_{\rm NIRSpec}$\tablenotemark{a} & $12.342\pm 0.009$ \\
$z_{\rm MIRI}$\tablenotemark{b} & $12.33\pm 0.04$ \\
$\nu_{\rm cent} $ [$\rm GHz$] & $254.487\pm 0.019$ \\
FWHM [$\rm km\,s^{-1}$] & $186 \pm58 $\\
$S_{\rm total}$ [$\rm mJy\,km\,s^{-1}$] & $36\pm10$ \\
$L_{\rm [OIII]}$ [$\rm L_\odot$]$\times10^8$ & $1.7\pm0.4$ \\ 
\enddata
\tablenotetext{a}{\citet{Castellano2024}} 
\tablenotetext{b}{\citet{Zavala2024}} 
%% Include any \tablenotetext{key}{text}, \tablerefs{ref list},
%% or \tablecomments{text} between the \enddata and 
%% \end{deluxetable} commands
%% No \tablecomments indicated
%% No \tablerefs indicated
\end{deluxetable}

\subsection{An upper limit on the [OIII]\,52$\mu$m line luminosity}\label{secc:upperlimit52um}
Armed with a more precise spectroscopic redshift from the [OIII] 88$\micron$ detection, we search for the 52$\mu$m line around the expected frequency of 433.96\,GHz, but no obvious line was detected (see Figure \ref{fig:52um_spectrum} in the Appendix). We estimate an upper limit on the [OIII] 52$\micron$ line luminosity using the measured local noise r.m.s., adopting the [OIII]88$\micron$ line-width of $186\pm58\,\rm km\,s^{-1}$ (see Table \ref{table:line_properties}) and assuming no spatially-extended emission beyond the beam-size. This results in a $3\sigma$ upper limit of $L_{\rm [OIII]52\mu m}<9.6\times10^8\,L_\odot$, implying a $3\sigma$ upper limit on the 52-to-88$\mu\rm m$ line ratio of $L_{\rm [OIII]52\mu m}/L_{\rm [OIII]88\mu m}<5.6$.

\section{Results}\label{secc:results}
\subsection{Detection of [OIII] 88$\,\mu\rm m$: Expectations and implications}\label{secc:scaling_relations}
\begin{figure*}[ht!]
    \centering 
    \includegraphics[width=0.6\textwidth]{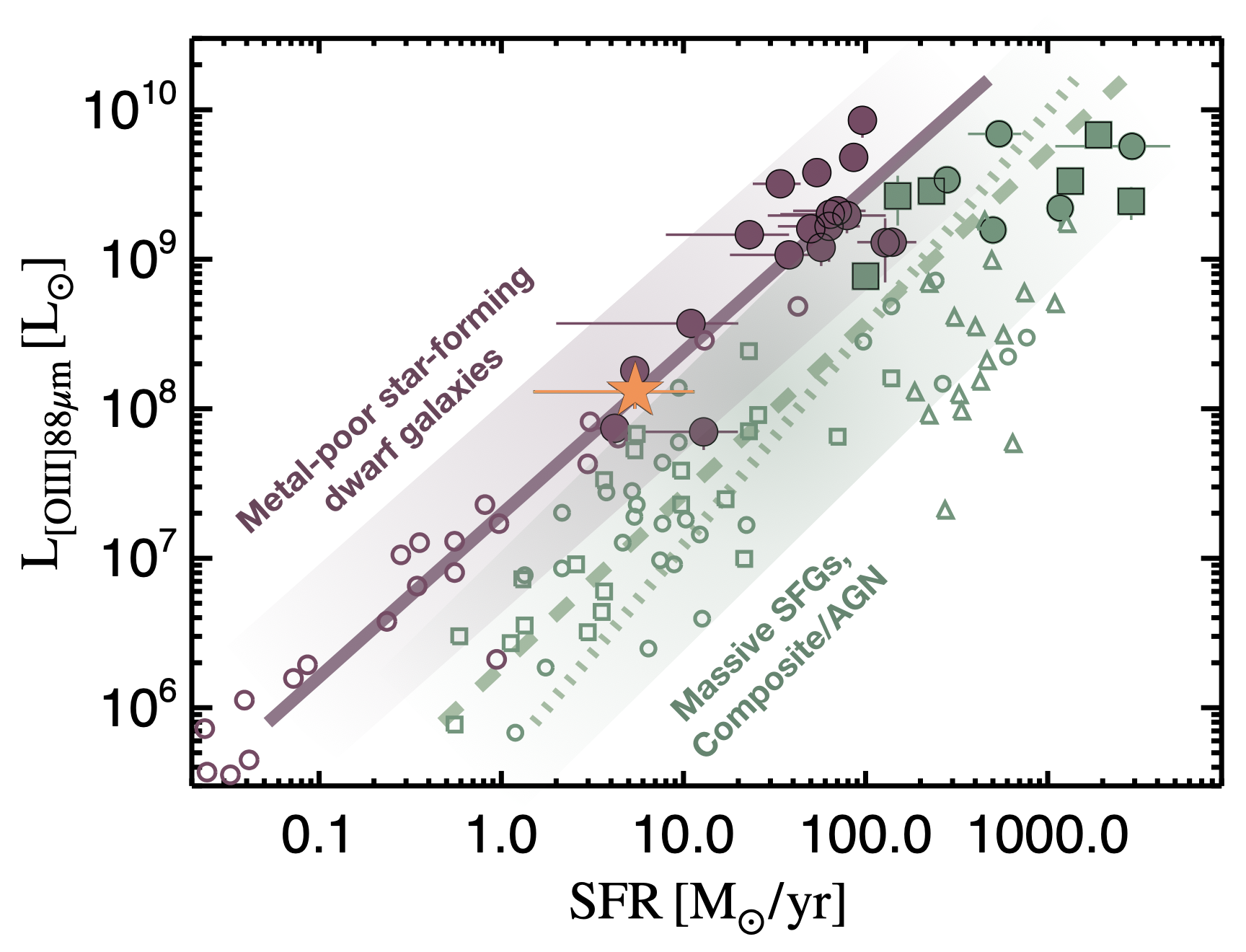}\vspace{-0.2cm}
    \caption{{\it The SFR-[OIII] line luminosity relationship.} The open symbols represent the sample of low-$z$ ($z\lesssim0.3$) galaxies presented in \cite{DeLooze2014}, with purple circles representing metal-poor "dwarf" galaxies, green open circles for more massive star-forming galaxies, green open squares for composite/AGN, and green open triangles for ultra luminous infrared galaxies. The relationships derived by \cite{DeLooze2014} are also plotted as solid, dashed, and dotted lines for the population of metal-poor "dwarf" galaxies, star-forming galaxies, and composite/AGN, respectively. Solid symbols represent high-z galaxies ($z\approx5-9$; as compiled in \citealt{Algera2024} and \citealt{Bakx2024}), with purple circles for LBGs/LAEs, green circles for submillimeter galaxies, and green squares for QSOs. Our target, GHZ2, is represented by the yellow star (adopting the range of SFRs from \citealt{Zavala2024} and correcting from magnification), and, along with other high-z LBGs, it nicely follows the local relationship for metal-poor "dwarf" galaxies. Note that the metallicity of GHZ2 is constrained to be $Z\approx0.05-0.1\,Z_\odot$ (\citealt{Calabro2024,Castellano2024,Zavala2024}). }
    \label{fig:sfr_vs_O3}
\end{figure*}

Using observations of local and low-redshift galaxies with the {\it Herschel Space Observatory}, \cite{DeLooze2014} found a set of correlations between the [OIII] 88$\micron$ line luminosity and galaxies' SFR for different galaxy populations. Evolved, massive star-forming galaxies (referred as {\it starbursts} in the aforementioned work) and composite/AGN systems are clearly offset from metal-poor star-forming dwarf galaxies. These correlations seems to be valid at $z\sim5-9$, as can be seen in Figure \ref{fig:sfr_vs_O3}, where we plot all the high-redshift [OIII] 88$\micron$ detections to date. Lyman break galaxies (LBGs) and Lyman-$\alpha$ emitters (LAEs) follow the same local relation found for metal-poor dwarf galaxies, while QSO and submillimeter-selected galaxies are in better agreement with the local relation for composite/AGN (although with large scatter and with some objects being consistent with both relationships). 

In this context, we show, in Figure \ref{fig:sfr_vs_O3}, the [OIII] 88$\micron$ line luminosity and SFR measured for our target and compare it with the reported relationships and with previous detections from high-redshift galaxies. We adopt the line luminosity reported above corrected by magnification ($\mu=1.3$; \citealt{Bergamini2023}), and the range of SFR values reported by \citealt{Zavala2024} based on the spectro-photometric SED fittings and the H$\alpha$ line ($2-12\,\rm M_\odot\,yr^{-1}$, after correcting for magnification). 

As can been seen in the figure, GHZ2 lies on a locus comparable to some $z\sim8$ LBGs/LAEs and in very good agreement with the local relationship for metal-poor dwarf galaxies. These  galaxies are dominated by star formation and, thus, by analogy, we suggest that it might also be the case for our $z=12$ target.  Actually, the measured line luminosity is also in broad agreement with the predictions from the {\sc serra} zoom-in simulations for galaxies with similar properties as GHZ2, in terms of SFRs and redshifts (\citealt{Kohandel2023}), which further support the star formation scenario for GHZ2. To further test this interpretation, below we compare other properties of GHZ2 with those measured directly for well-known systems dominated by star formation, such as HII regions and HII galaxies\footnote{HII galaxies are compact low-mass systems ($M_\star<10^9\,M_\odot$) whose luminosities are almost completely dominated by a massive recent burst of star formation; see \citet{FernandezArenas2018}; \citet{Gonzalez-Moran2021}; and reference therein.} (while the potential contribution of an AGN in discussed in Section \ref{secc:discussion}).

\begin{figure*}[ht!]
    \centering \includegraphics[width=0.61\textwidth]{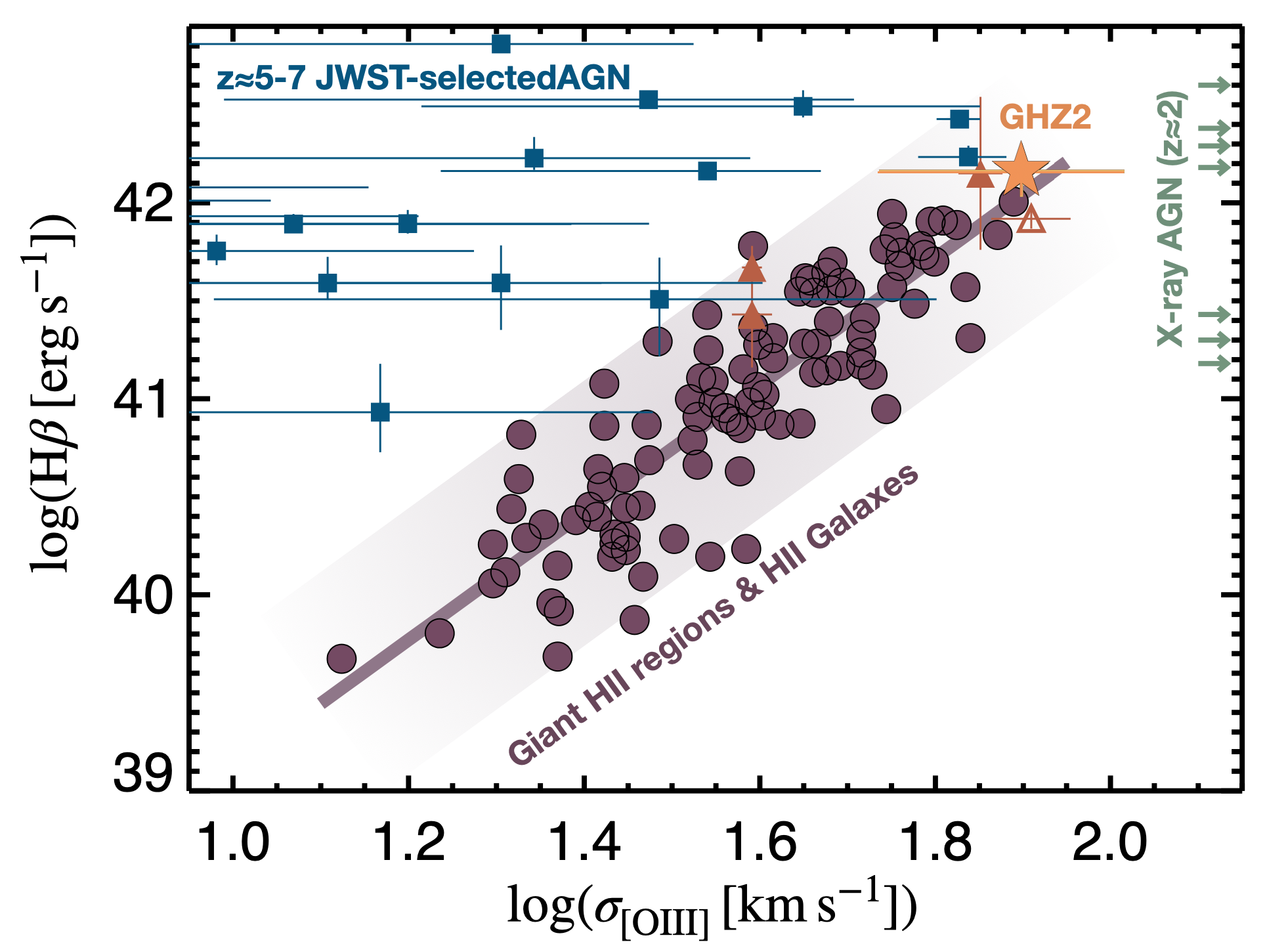}
    \caption{{\it The $\rm H\beta$ line luminosity-velocity dispersion relationship. } The position of GHZ2 in the $L-\sigma$ relationship for Giant HII regions and HII-galaxies is represented by the yellow star (adopting the lensing-corrected $L_{\rm H\beta}$ reported in \citealt{Zavala2024} and the line-width from the ALMA [OIII] 88$\micron$ detection reported in this work). As can be seen, the $z=12$ galaxy follows the relation found for HII-galaxies (as reported by \citealt{Melnick2017}; solid purple line and shaded region), suggesting a similar nature for both objects: young burst of star formation. The three orange solid triangles show $z\approx6-7$ HII-galaxies recently identified reported (\citealt{Chavez2024}) while the open orange triangle represents MACS0647‑JD at $z=10.2$ (\citealt{Hsiao2024}). Finally, we plot the high-$z$ AGN sample from \citet{Taylor2024} and indicate the position of more massive AGN at $z\sim2$ (\citealt{Kakkad2020}), with both populations deviating from the relationship for HII-dominated systems.}
    \label{fig:Lsigma}
\end{figure*}

Over the past decades, it has been shown that giant HII regions exhibit a correlation between the luminosity of the Balmer emission lines (typically $H\beta$) and the line velocity dispersion (\citealt{Terlevich1981}). Although the physical origin of this relation is still a matter of debate,  it is typically thought to be driven by the total mass of  the (gravitationally-bounded) star-forming region, dominated by young super star clusters. The relation has been calibrated using samples of several dozens of local  and extragalactic giant HII regions (e.g. \citealt{Chavez2014}) and shown to be also valid for  HII-galaxies up to $z\sim3$ (e.g. \citealt{Gonzalez-Moran2021}).  \citet{Melnick2017} presented a new calibration using the $\rm [OIII]5007\AA$ line-width and the $H\beta$ line luminosity that shows a similar scatter than the original $L_{\rm H\beta}-\sigma_{\rm H\beta}$ relationship. Here we use this correlation given that the Balmer lines are not spectrally resolved for our target, and test whether GHZ2 follows a similar trend, as expected if the source is dominated by  young bursts of star formation and assuming the relationship does not evolve significantly with redshift (note that the $z\sim6-7$ HII-galaxies recently reported by \citealt{Chavez2024} follow the same relationship, as can be seen in Figure \ref{fig:Lsigma}).

We adopt the lensing-corrected H$\beta$ luminosity of $\log\, L_{\rm H\beta} = 42.16^{+0.09}_{-0.13}\rm\,erg\,s^{-1}$ from \citet{Zavala2024} (inferred from the H$\alpha$ line assuming zero dust attenuation), and the [OIII] 88$\micron$ line-width provided by the ALMA observations as a proxy for the [OIII]5007$\AA$ velocity dispersion (given that both transitions are produced by the same ion and the [OIII]5007$\AA$ line is unresolved in the MIRI low-resolution spectrum). As can be seen in Figure \ref{fig:Lsigma}, GHZ2 follows the same relationship found for giant HII regions, suggesting that this very high-redshift galaxy is also powered by (extreme) star formation. Actually, some of the [OIII] 88\micron-detected dwarf galaxies described above are also known to harbor bright HII regions powered by super star clusters (\citealt{Madden2013}). 
This might imply that the high ionization conditions in GHZ2 are likely driven by extreme HII regions associated with young bursts of star formation. The same can be said for the $z=10.2$ MACS0647-JD galaxy recently observed with NIRSpec and MIRI (\citealt{Hsiao2024,Hsiao2024b}). This object also follows the relationship derived for HII-dominated systems, which  supports the conclusions by \cite{Hsiao2024b} who argued against an AGN-dominated galaxy.

On the other hand, while similar relationships between the Balmer line luminosities and lines' velocity dispersion have been reported for AGN (e.g. \citealt{Ho2003}), typical AGN hosting super massive black holes show emission lines significantly broader than the HII regions and HII galaxies (with [OIII] line-widths of $>300\,\rm km\,s^{-1}$; e.g. \citealt{Kakkad2020}). And while less massive AGN with narrower line-widths have recently been found by JWST up to $z\sim7$ or beyond, they have brighter line luminosities compared to star-forming dominated systems (see Figure \ref{fig:Lsigma}). All of this supports our conclusion that GHZ2 is most likely dominated by a burst of star formation.

Given that the population of young, low-metallicity massive clusters powering giant HII regions have been proposed to be the progenitors of globular clusters (e.g. \citealt{Ho1997,PortegiesZwart2010,Terlevich2018}), it is possible that the current starburst episode of GHZ2 might result in the formation of proto-globular clusters or other compact stellar systems. Indeed, its abundance, with an enhanced nitrogen abundance, has already been  noticed to be consistent with the expected progenitor population for globular clusters (\citealt{Castellano2024}), as well as its high SFR and stellar mass surface densities (\citealt{Ono2023,Calabro2024}). This might point towards an evolutionary connection between the bright, compact high-redshift galaxies discovered by JWST  (including GHZ2) and the enigmatic population of old globular clusters (see also \citealt{Charbonnel2023,Isobe2023,Castellano2024,Marques-Chaves2024,Topping2024}). This is discussed in detail below,  in Section \ref{secc:discussion}.

\subsection{Dynamical mass}\label{secc:mdyn}

Assuming a spherical geometry (a reasonable assumption for this very compact source), we estimate the dynamical mass using the virial equation (e.g. \citealt{Pettini2001,Wolf2010}):
\begin{equation}
     M_{\rm dyn}=\frac{5\sigma^2 R}{G}, \label{eq:Mdyn}
\end{equation}
where $\sigma$ is the velocity dispersion and $R$ is the virial radius. The velocity dispersion is determined from the [OIII] 88$\micron$ transition ($\sigma=79\pm25\rm\,km\,s^{-1}$) and the virial radius is adopted to be the  half-light radius measured from the JWST/NIRcam images, with reported values of $r_{1/2}=39\pm11\,$pc (\citealt{Ono2023}) and $105\pm9\,$pc (\citealt{Yang2022}). Using Equation \ref{eq:Mdyn} and the aforementioned assumptions we constrain the dynamical mass to be  $3\times10^8\rm-8\times10^8\rm\,M_\odot$. 

This is remarkably similar to the stellar masses inferred from the spectro-photometric SED fitting analysis presented in \citet{Zavala2024}, with  best-fit stellar masses of $2\times10^8\rm-8\times10^8\rm\,M_\odot$ (assuming a magnification of $\mu=1.3$: \citealt{Bergamini2023}). This implies a dynamical-to-stellar mass ratio of close to unity for GHZ2, similar to the measured values for globular clusters (e.g. \citealt{Forbes2014}), providing more evidence for a relation between these objects as discussed below.

Finally, to assess the impact of our assumptions on the inferred dynamical mass, we alternatively use the approach described by \citealt{Ubler2023}. Here, we assume Sersic indices of $n=1$ and $1.5$, with axis ratios in the range of $q=\{0.3:0.7\}$. Overall, the dynamical masses inferred from these models fall within the range of the values described above but allow for up to a 50\% increase. However,  given the compactness and symmetry of the source (\citealt{Treu2023b}), we adopt the mass constraints from the virial equation described above as our fiducial value.

\subsection{Constraining the electron density}\label{secc:electron_density}

\begin{figure}[h]   \hspace{-0.5cm} \includegraphics[width=0.51\textwidth]{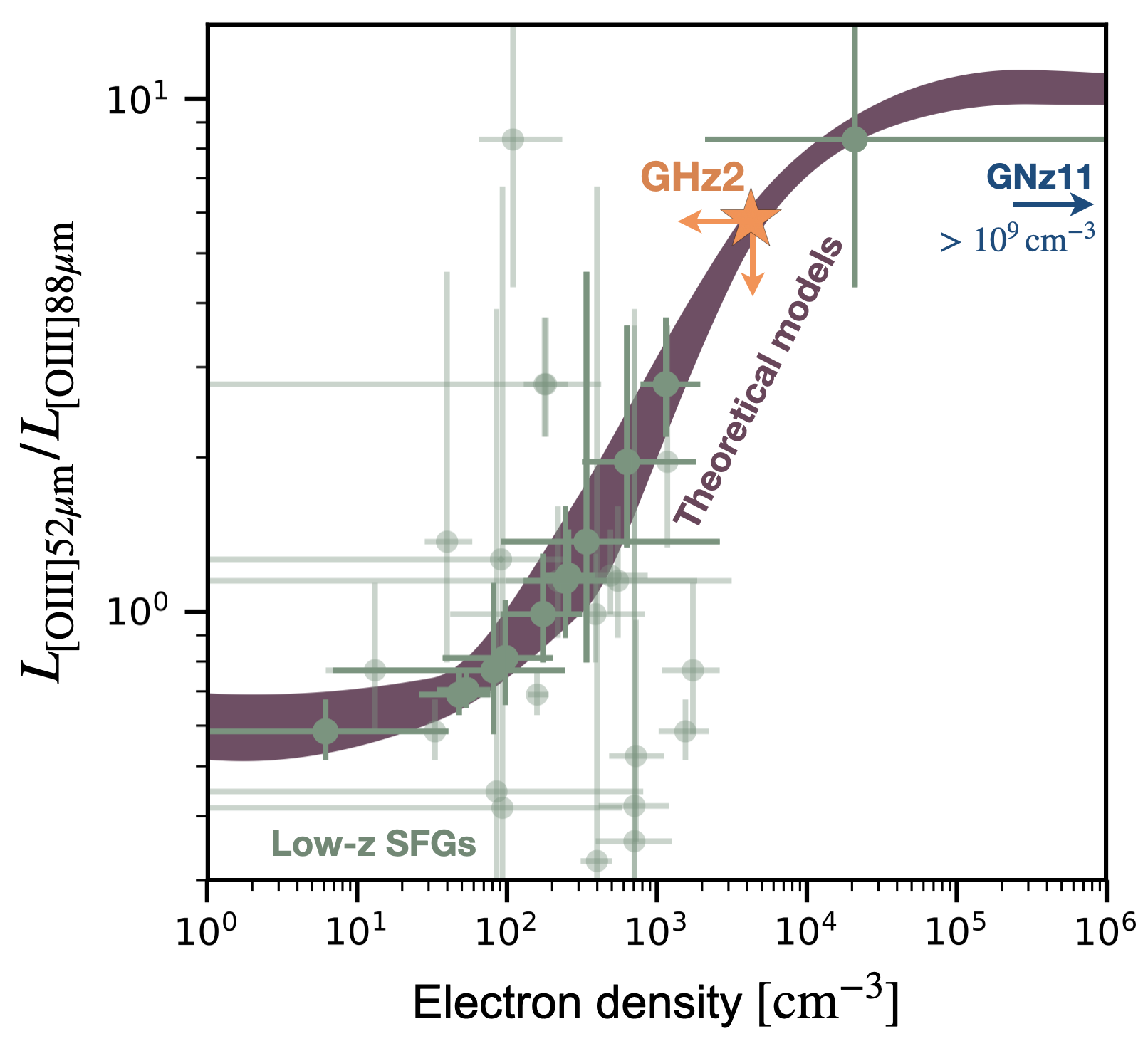}
    \caption{{\it Constraints on the Oxygen electron density.} Predicted 52-to-88~$\mu$m luminosity ratio as a function of electron density generated using {\sc pyneb} (\citealt{Luridiana2015}). This ratio is nearly-independent to gas temperature, ionization parameter and metallicity. The 3$\sigma$ upper limit from our observations (orange star) rules out high-density environments, such as those found in GNz-11 ($>10^9\,\rm cm^{-3}$). The comparison data come from the PACS and SPIRE spectrometry of nearby star-forming galaxies (\citealt{Fernandez-Ontiveros2016}). The solid green symbols indicate the Oxygen-derived electron densities, which accurately follow the theoretical predictions, while the semi-transparent symbols indicate complementary probes of the electron densities (using [NII]205/122, [SIII]33/18, and [NeV]24/14 ratios). As can be seen, star-forming galaxies show moderate electron densities, similar to those inferred for GHZ2.} 
    \label{fig:n_e}
\end{figure}

The [OIII]52$\micron$-over-[OIII]88$\micron$ line luminosity ratio provides strong constraints on the electron density almost independently from metallicity, ionization field effects, or electron temperature (e.g. \citealt{Jones2020,Killi2023,Nakazato2023}). Here, given the non-detection of the [OIII] 52$\mu$m transition (see Section \ref{secc:upperlimit52um}), we determine a $3\sigma$  upper limit on the ionized Oxygen electron density of $\log(n_e\,[\rm cm^{-3}])<3.62$ (see Figure \ref{fig:n_e}). While this value seems high compared to the typical values found in star-forming galaxies at low redshifts ($\sim100\,\rm cm^{-3}$; e.g. \citealt{Kaasinen2017}), this upper limit is totally consistent with the measured electron densities of high-redshift ($z\gtrsim5$) star-forming galaxies and with the apparent  redshift evolution of $n_e$ (e.g. \citealt{Isobe2023,Jones2024}), particularly when combined with the lower limit of $n_e\gtrsim100\,\rm cm^{-3}$ reported in \citet{Zavala2024}. On the other hand, the electron density of GHZ2 contrasts with that of GNz11, the brightest confirmed galaxy at $z>10$. In the case of GNz11, the  [NIV] doublet and NIII] multiplet  indicate extremely large electron densities ($>10^5\rm cm^{-3}$) in better agreement with the broad line regions in AGN (\citealt{Maiolino2024}). We stress, however, that transitions with different ionization potentials might trace different regions of the ISM with different electron densities (e.g. \citealt{ji2024})

\section{Discussion}\label{secc:discussion}
\subsection{A system dominated by star formation}\label{secc:SSC}
The nature of the intriguing population of bright, high-redshift galaxies discovered by JWST remains unclear. {\it What processes drive their high luminosity so early in the Universe's history?}
Understanding whether these galaxies are dominated by stellar processes or AGN activity is crucial for developing a comprehensive picture of galaxy formation in the early Universe.

In the case of GHZ2, this question has been approached from different perspectives, starting with its JWST/NIRCam morphology.  \citet{Ono2023} found that while the source is very compact, it is spatially resolved with a half-light radius of a few tens of parsecs (see also \citealt{Yang2022}). 
This provides evidence against an AGN-dominated system, although it does not totally rule out some fractional contribution to its continuum luminosity. Similarly, \citealt{Zavala2024}, who presented the detection of $H\alpha$ and [OIII]5007$\AA$ in the JWST/MIRI spectrum and a spectro-photometric SED analysis of this source, attributed its high luminosity to the presence of a very young, massive and metal-poor stellar population; but, again, the possibility of having some contribution from AGN remained open. This latter scenario was revitalized given the detection of relatively high-ionization UV transitions and line rations, and lines with high equivalent widths, in the JWST/NIRSpec spectrum presented by \citet{Castellano2024}, indicating ionizing photons with energies in excess of $\sim 50-60$\,eV. Nevertheless, the absence of higher ionization lines typically seen in AGN (like HeII or [NeIV]2424$\AA$) and the strong  upper limit on the [Ne IV]/N IV] ratio put some doubts on this scenario and led the authors to suggest that GHZ2 is most likely dominated by (extreme) star formation. Finally, \citet{Calabro2024}  presented a comprehensive analysis of the (rest-frame) UV and optical line ratios finding very high ionization conditions but indistinguishable between those produced by AGN and low metallicity, compact star-forming regions.

It is thus clear that GHZ2 shows extreme properties  with high ionization conditions, but a pure AGN model would fail to reproduce all the characteristics described above (such as the lack of some specific atomic transitions and extended emission). Additionally, in Section \ref{secc:scaling_relations} we have shown that this source follows the SFR-[OIII] 88$\micron$ relationship found for metal-poor star-forming galaxies  and the L-$\sigma$ relation found for giant HII regions and HII-galaxies, which would be hard to explain within the AGN framework. The fact that the H$\alpha$-based SFR is in good agreement (within a factor of $\sim2$) with the star-forming SED modelling (see \citealt{Zavala2024}) might also imply that the contribution from AGN (if any) should be low. The same could be said of the fact that the stellar mass (from the SED analysis) and the dynamical mass (from the [OIII] 88$\micron$ detection) are consistent with a factor of $\sim 2$. Such results would be entirely coincidental in the case of an AGN-dominated system. Finally, the electron density upper limit of $n_e<4.2\times10^3\,\rm cm^{-3}$, although high, has been shown to be within the realm of star-forming galaxies (see \S\ref{secc:electron_density}), and significantly below the typical values of the broad line region in AGN. Hence, we conclude that GHZ2 is most likely dominated by star formation activity driven by metal-poor and young stellar populations.  

Furthermore, given its high H$\alpha$ and H$\beta$ luminosities -- consistent only with the most luminous starbursts and HII galaxies (e.g. \citealt{Ho1997b,FernandezArenas2018}) -- and its remarkable compactness and concentration (\citealt{Ono2023,Treu2023b}), it would be expected that this extreme star formation activity is taking place in the form of super star clusters, commonly found in starburst systems (\citealt{Ho1997b}).  The high specific SFR of GHZ2 of $\rm sSFR\approx10-30\,Gyr^{-1}$ might actually be an evidence of a starburst-like star formation mode propitious to the formation of these massive clusters. This is also supported by the high {\it burstiness} found for the {\sc serra} simulated galaxies that match other properties of GHZ2 (\citealt{Pallottini2022}).

However, it is important to remark that, although the star-forming scenario seems the most plausible to explain the nature of this object, available models fail to reproduce the extremely high equivalent widths of the UV lines, some of them  only compatible with AGN or composite emission models, as discussed in detail by \citet{Castellano2024} and \citet{Calabro2024}.

\begin{figure*}[ht!]
    \centering 
    \includegraphics[width=1.02\textwidth]{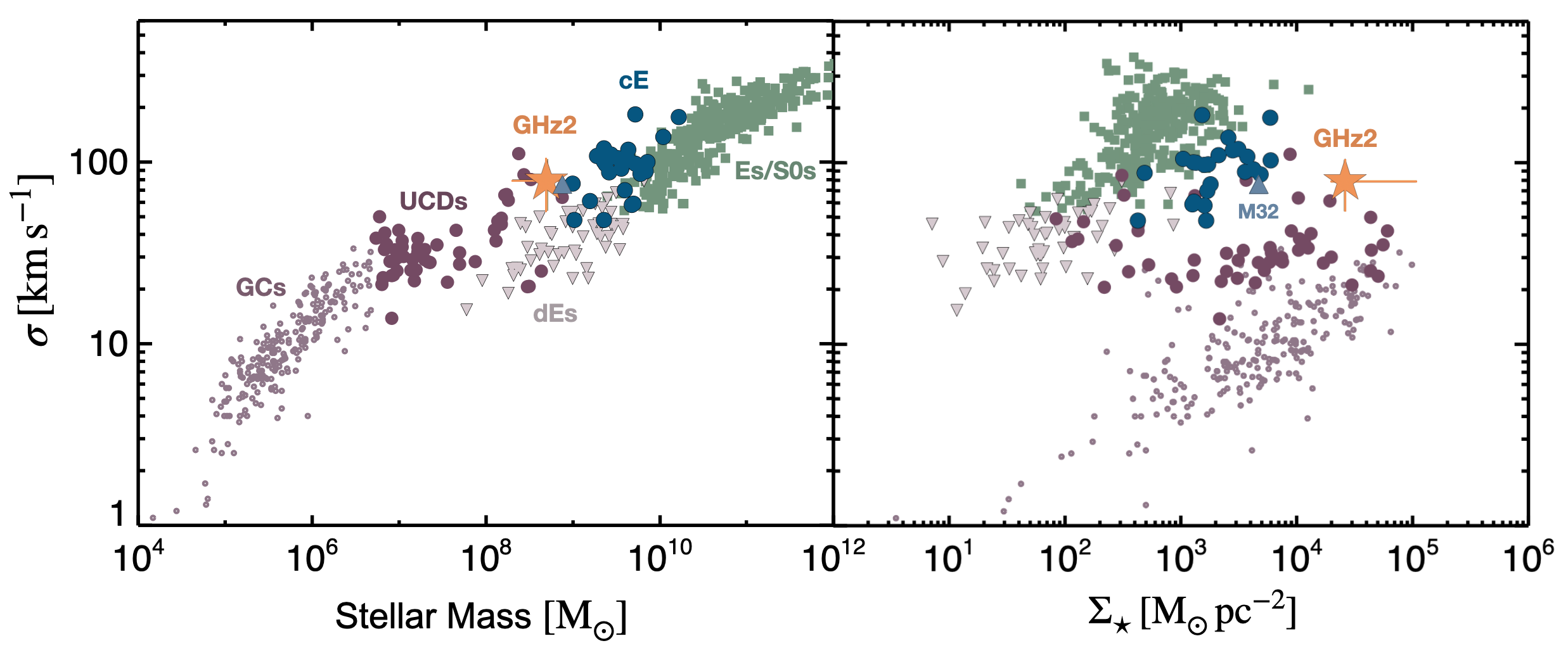}
    \caption{{\it The stellar mass-velocity dispersion relationship.} The velocity dispersion as a function of stellar mass (left) and stellar mass surface density (right) for different populations of pressure-supported systems (taken from \citealt{Norris2014}), including globular clusters (light purple circles), ultra-compact dwarfs (solid purple circles), dwarf ellipticals (down-pointing triangles), compact ellipticals (blue solid circles), and ellipitcal and S0 galaxies (green squares).  Our target, GHZ2, represented by the yellow star, shows a similar stellar mass surface density as globular clusters, but with a higher total stellar mass that is in better agreement with those of ultra-compact dwarfs and compact ellipticals. Indeed, it occupies a  similar locus as M32, the prototypical compact elliptical galaxy (represented by the up-pointing blue triangle). Note, however, that significant mass losses are expected due to stellar and dynamical evolution that would eventually affect the position of GHZ2 on this parameter space.  
    } 
    \label{fig:mstar-sigma}
\end{figure*}

\subsection{The progenitor of a compact stellar object}
{\it Will GHZ2 evolve into a globular cluster?}
%Yes, but...
The fact that GHZ2 might be powered by massive  star clusters does not necessarily make it a globular cluster progenitor. However, here we provide additional arguments that might support this scenario and that might, additionally,  explains some of the long-standing problems associated with globular clusters.

First, its low metallicity and enhanced nitrogen abundance (\citealt{Castellano2024}) resembles the well-known chemical anomalies observed in globular clusters (e.g. \citealt{Cannon1998}) -- though not exclusive to them. These abundance patterns are thought to be the product of massive stars in very dense environments (e.g. \citealt{Smith2006}) and, interestingly, in GHZ2, the presence of very massive stars has been inferred from its high ionization conditions (\citealt{Calabro2024,Castellano2024,Zavala2024}). Moreover, its compact morphology implies a high  mass  density (with $\Sigma_\star\sim10^4\,\rm M_\odot\,pc^{-2}$ and $n_e\approx 100-4,000\,\rm cm^{-3}$), similar to those found in globular clusters (see Figure \ref{fig:mstar-sigma}). 
In addition, spectro-phometric SED fitting analysis on GHZ2 indicates a bursty star formation history for this object (\citealt{Zavala2024}; see also \citealt{Harikane2024b}). These short bursts with duration in the range of a few or a few tens of Myr, in combination with its very rapid  metal enrichment\footnote{Note that, despite its young age, GHZ2 shows already a metallicity of $Z\approx5-10\%\,Z_\odot$; \citealt{Calabro2024,Castellano2024,Zavala2024}.}, would naturally explain the low spread in stellar ages and the multiple stellar populations observed in globular clusters (see reviews by \citealt{Gratton2012}, \citealt{Forbes2018}). Lastly, its unique spectral features with UV lines with high EWs and clear emission of the OIII\,3123\AA\, fluorescence line\footnote{This transition is pumped by photons emitted from singly ionized helium atoms; i.e. He\,II Ly$\alpha$ photons at $\lambda_{\rm rest}=303.78\,\AA$.} (\citealt{Castellano2024}) might be explained by the {\it stellar exotica} in globular clusters (e.g. X-ray binaries; \citealt{PortegiesZwart2010}) and their high helium abundance (\citealt{Norris2004,DAntona2005}).

There is one thing, however, that appears to be at odds with expected proto-globular cluster properties: the high total mass of GHZ2. With a stellar mass of $\approx10^8-10^9\,\rm M_\odot$ (see Section  \ref{secc:mdyn}), GHZ2 exceeds by more than one order of magnitude the masses of even the most massive known globular clusters (see Figure \ref{fig:mstar-sigma}). These two values might be reconciled by assuming significant mass losses  due to stellar and dynamical evolution (e.g. \citealt{Webb2015}), but even assuming extreme values, GHZ2 would only be barely consistent with the most massive population of globular clusters. 
Alternatively, this tension might be solved by the models presented in \citet{DErcole2008}, in which most of the first-generation stars are lost due to the expansion and stripping of the cluster's outer layers caused by the heating from SN ejecta, followed by a highly concentrated second generation of stars.  Indeed, several models require this first generation to be at least $\sim10$ times more massive than current globular cluster survivors in order to explain the different chemical composition of the multiple stellar populations (e.g. \citealt{Renzini2015}).

Another possibility is that GHZ2 hosts an ensemble of super star clusters. This scenario is supported not only by its large mass, but also by its size. Its effective radius of  $\sim40\,\rm pc$ exceeds by a factor from a few to tens the sizes of the few high-redshift proto-globular clusters that have been individually resolved thanks to strong gravitational lensing effects (e.g. \citealt{Adamo2024,Vanzella2023}). These two discrepancies (the size and total mass), could be solved by assuming that GHZ2 is actually a superposition of several individual clusters (e.g. \citealt{Mowla2024}). Supporting evidence for this scenario is found in the simulations presented in \citet{Kohandel2023}, in which the [OIII] 88$\mu$m emission of $z>10$ galaxies is resolved into several individual HII regions. And, although it is possible that these HII regions might be normal OB associations with a high probability of dissolution, as discussed above and in Section \ref{secc:SSC}, GHZ2 shows properties rather similar to those of massive, super star clusters.

The final scenario discussed considers the possibility that GHZ2 is an ultra-compact dwarf (UCD) galaxy or a proto-compact elliptical  given the similarities between their stellar mass, size, and velocity dispersion (see Figure \ref{fig:mstar-sigma}). 
The formation processes of these objects remains unclear, but they are known to bridge the gap between stellar clusters and  galaxies (e.g. \citealt{Norris2014,Terlevich2018}; see also Figure \ref{fig:mstar-sigma}), and it has been shown that some of these objects could represent the massive end of the globular cluster luminosity function (e.g. \citealt{Mieske2012}) or the result of merging of several star clusters (e.g. \citealt{UrrutiaZapata2019}). While it is also true that some UCD show elevated dynamical-to-stellar mass ratio that can be attributed to the presence of massive black holes (e.g. \citealt{Seth2014}), GHZ2 shows a dynamical-to-stellar mass ratio around one (see Section \ref{secc:mdyn}), in better agreement with globular clusters and compact ellipticals (\citealt{Forbes2014}).

We thus conclude that GHZ2 might be powered by young and compact star-forming regions that could evolve into globular clusters or to become the central core of a more massive galaxy. This is, however, only suggestive and not conclusive. Further observations and simulations should be conducted to further test this interpretation.

\vspace{1cm}
\section{Summary}\label{secc:summary}
Through the detection of [OIII] 88$\micron$ at $z=12.33$, we have demonstrated the feasibility of studying the early ($z>10$) universe with ALMA and its unique synergy with JWST to characterize the most distant galaxies known to date. By combining these observations with previous multi-wavelength photometry and spectroscopy, we investigated the nature of GHZ2 -- one of the brightest known galaxies at $z>10$ universe, just 350 million years after the Big Bang. We concluded that this object is likely powered by young bursts of star formation similar to those found in giant HII regions (metal-poor, massive, and compact star clusters).  Finally, we  presented compelling evidence suggesting that GHZ2 might be a progenitor of globular clusters or represent the early phases of a compact dwarf or elliptical galaxy. While this characterization is very suggestive, further observations with both high spectral and high angular resolution are needed to test this hypothesis and to fully understand some of the unique properties of this object.

\begin{acknowledgments}
We thank Stefano Carniani, Mahsa Kohandel, and Andrea Ferrara for stimulating discussions; Angela Adamo,  Hiddo Algera, Ricardo Ch\'avez, and Anthony Taylor for sharing their datasets; and Takuya Hashimoto and Yurina Nakazato for sharing information about their research projects. We also thank the anonymous reviewer and the scientific editor for their valuable feedback on the manuscript. JAZ acknowledge funding from JSPS KAKENHI grant number KG23K13150.  MC and PS acknowledge INAF Mini Grant 2022 “Reionization and Fundamental Cosmology with High-Redshift Galaxies, INAF Mini Grant 2022 “The evolution of passive galaxies through cosmic time”,  and PRIN 2022 MUR project 2022CB3PJ3 - First Light And Galaxy aSsembly (FLAGS) funded by the European Union - Next Generation EU. LCH was supported by the National Science Foundation of China (11991052, 12233001), the National Key R\&D Program of China (2022YFF0503401), and the China Manned Space Project (CMS-CSST-2021-A04, CMS-CSST-2021-A06). PGP-G acknowledges support from grant PID2022-139567NB-I00 funded by Spanish Ministerio de Ciencia, Innovaci\'on y Universidades MICIU/AEI/10.13039/501100011033, and the European Union FEDER program {\it Una manera de hacer Europa}.
This paper makes use of the following ALMA data: ADS/JAO.ALMA\#2021.A.00020.S and JAO.ALMA\#2023.A.00017.S. ALMA is a partnership of ESO (representing its member states), NSF (USA), and NINS (Japan), together with NRC (Canada), MOST,ASIAA (Taiwan), and KASI (Republic of Korea), in cooperation with the Republic of Chile. The Joint ALMA Observatory is operated by ESO, AUI/NRAO, and NAOJ. 
\end{acknowledgments}

\facilities{ALMA (Band 6 and Band 8; \citealt{Ediss2004,Sekimoto2008}), {\it JWST} (\citealt{Gardner2006,Gardner2023})}


\software{CASA \citep{CASATeam2022},  
          }



\appendix

% \section{A note on the previous ALMA non-detections and future follow-ups on high-redshift galaxies}

% $\bullet$ Highly uncertain redshifts (or even interlopers) $->$ spectral scan (i.e. blind surveys) vs targeted lines. \\
% $\bullet$ WSU 

\section{Band 8 spectrum around the [OIII] $52\,\mu \mathrm{\lowercase{m}}$ transition}
\begin{figure*}[h]
    \centering
    \includegraphics[width=\textwidth]{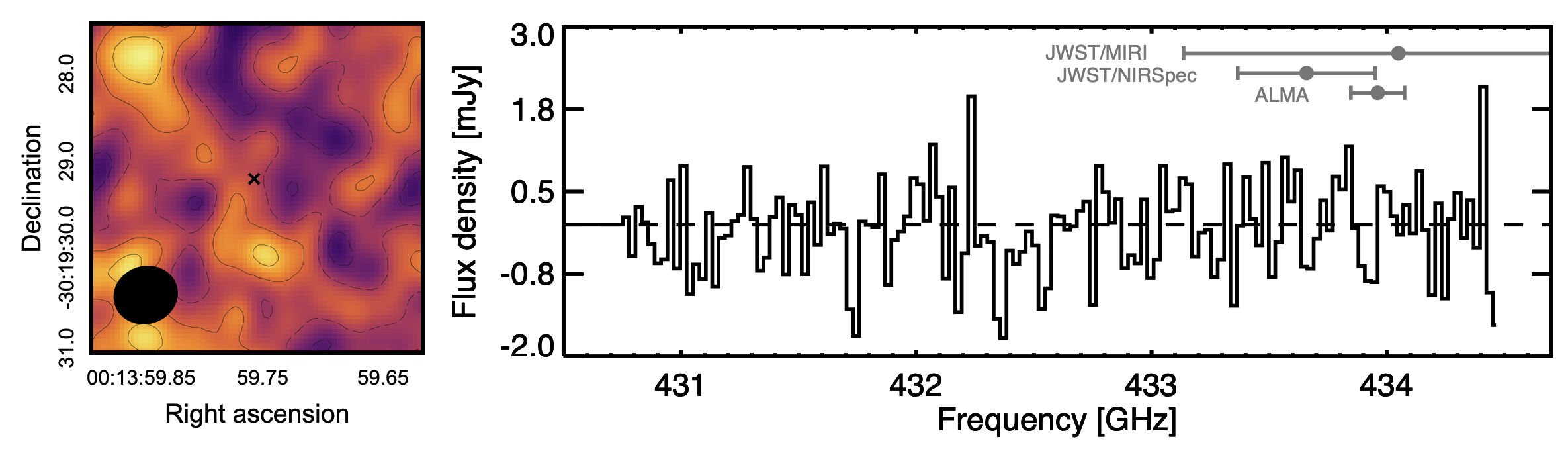}
    \caption{{\it ALMA Band-8 spectrum covering the [OIII]\,52\,$\mu\rm m$ transition at $z\approx12.3$}. Analog to Figure 1, but for the 52\,$\mu\rm m$ observations. The left panel shows a $3\farcs5\times3\farcs5$ image of the collapsed datacube within the expected FWHM of the line (based on the [OIII]\,88\,$\mu\rm m$ detection). Nevertheless, there is no significant detection around the position of GHZ2 (black cross). Contours show $\pm1\sigma$ and $\pm2\sigma$, with dashed lines for negative values. The extracted spectrum at the position of the source is shown on the right panel along with the redshift constraints from the different instruments. }
    \label{fig:52um_spectrum}
\end{figure*}

Figure \ref{fig:52um_spectrum} shows the extracted spectrum at the position of GHZ2 (right panel) along with a collapsed data-cube within the expected FWHM of the line (from approx. 433.8 to 434.1\,GHz; left panel). The line is undetected at the current depth of the observations, but provide useful constraints on the electron density as discussed in Section \ref{secc:electron_density}.

\bibliography{sample631}{}
\bibliographystyle{aasjournal}



\end{document}